\documentclass{article}
\usepackage{amsmath,amssymb,epsfig}

\newcommand{\ket}[1]{| #1 \rangle}
\newcommand{\bra}[1]{\langle #1 |}
\newcommand{\mean}[1]{\langle #1 \rangle}
\newcommand{\atan}{\operatorname{atan}}
\begin{document}

\title{Sensitivity to Initial Conditions in Quantum Dynamics:
an Analytical Semiclassical Expansion}
\author{K.M. {Fonseca Romero}$^{(1)}$, 
M.C. Nemes$^{(2,3\footnote{Permanent
Address.}\;\;)}$,\\ 
J.G. {Peixoto de Faria}$^{(2)}$, 
and A. F. R. de Toledo Piza$^{(2)}$}

\maketitle

\begin{center}

{$^{(1)}$ Departamento de F\'{\i}sica, Universidad Nacional,
Bogot\'a, Colombia}

{$^{(2)}$ Departamento de F\'{\i}sica--Matem\'atica, Instituto de
 F\'\i sica, Universidade de S\~ao Paulo,\\  C.P. 66318, 05315-970
 S\~ao Paulo, S.P., Brazil}

{$^{(3)}$ Departamento de F\'\i sica, ICEX, Universidade Federal de
 Minas Gerais,\\ C.P. 702, 30161-970 Belo Horizonte,
 M.G., Brazil}

\end{center}

\begin{abstract}

We construct a
class of systems for which quantum dynamics can be expanded around
a mean field approximation with essentially classical content.
 The modulus of the quantum overlap of mean field states 
naturally introduces a classical distance between classical phase 
points. Using this fact 
we analytically show that the time rate of change (trc) of two
 neighbouring classical trajectories is directly proportional to the 
trc of quantum correlations. Coherence 
loss and nonlocality effects appear as corrections to mean
field dynamics and we show that they can be given in terms of 
classical trajectories and
generalized actions. This result is a first step in the 
connection between quantum and classically chaotic dynamics in the
same sense 
of semiclassical expansions for the 
density of states. We apply the results to the nonintegrable
(classically chaotic) version of the N-atom Jaynes-Cummings model.

\end{abstract}




Ever since the conception of Quantum Mechanics the classical limit
has been a matter of much debate due to the profound
contrasting differences between the classical and quantal descriptions
of the world. Although the difficulties in building a bridge between
quantum and classical mechanics are well known, we start by reviewing
 the ones which are of relevance for the present contribution.
As far as kinematical differences  are concerned,
already at the level of a point particle, striking differences 
appear. While
the definition of the state of a classical particle 
is of local character and given by a point in phase space,
the quantum counterpart of the definition of a particle state is
given by a vector in Hilbert space which cannot simultaneously
be ascribed a well defined value for position and momentum.
The closer one can get to the classical situation are minimum
uncertainty wave packets. Quantum states are therefore usually
nonlocal. Also, the linear character of the Hilbert space has
the immediate consequence that superposition of (minimum
uncertainty) states
are also possible states and, in fact, constitute the vast majority of
allowed quantum states. The situation gets even cloudier
when two degrees of freedom are involved: Classically one can always
describe a two particle state in terms of the coordinates and 
momenta of each one of them. Quantum mechanically, however,
this situation only holds if the two particles, initially in a 
factorized state, do not interact.
The Hilbert space structure allows for states which cannot be
written as a direct product of vectors in the individual Hilbert
spaces of each degree of freedom. This essentially quantum property 
is usually named entanglement. Much investigation and progress both
on the theoretical and experimental sides have been achieved recently.
\cite{entanglement}

From the dynamical point of view one of the essential differences 
has given rise to an important research area nowadays: classical chaos,
a phenomenon whose root lies on the nonlinearity of Newton's equation.
The relevant question here is how to identify the quantum counterpart
of classical chaos. A major step in this direction was given by
Bohigas and collaborators who conjectured that spectral properties 
of integrable and nonintegrable systems should be very different 
\cite{Bohigas}.
Thereafter many numerical investigations confirmed such conjecture
and a few exceptions where found. From the analytical point of view a 
most relevant formula was derived by Gutzwiller connecting level
densities of very general quantum systems with classical periodic
orbits and their actions\cite{Gutzwiller}. 
In what concerns the connections between
classical and quantum dynamics it has recently been proposed by 
Zurek that the rate of entropy 
production can be used as an intrinsically quantum test of the 
chaotic vs. regular nature of the evolution\cite{Zurek}.
Several numerical tests of this conjecture can also be 
found\cite{Kyoko,Miller}.
Anyway, analytic results in this context are scarce.
This letter is a step in the direction of filling in this gap.

Given the considerations above we restrict ourselves to the study
of the (two degrees of freedom) hermitian bilinear hamiltonians
\begin{equation}\label{bihamiltonian}
H=\sum_i \alpha_i A_i
 +\sum_j \beta_j B_j
 +\sum_{i,j} \gamma_{i,j} A_i B_j,
\end{equation}
where $i,j = {0,\pm}$, A and B are chosen among the generators 
of either the $h(3)$ (Heisenberg group)
or $SU(2)$ groups.
Index 0 is associated with the operator $a^\dagger a$
($J_z$), the index $+$ with the operator $a^\dagger$ ($J_+$) and
the index $-$ with the operator $a$ ($J_-$) for the algebra
$h(3)$ ($su(2)$). The coefficients $\alpha$, $\beta$ and $\gamma$ 
are constants or given functions of time.

Although seemingly trivial this class of systems encompasses a rich
variety of dynamical behavior including the model we shall use for
illustration whose clasical limit is chaotic. The choice of the 
groups $h(3)$ and $SU(2)$ is due to the fact
that several of the problems mentioned above can be circumvented at 
the lowest order, the mean field. Due to the bilinear character
of the hamiltonian, a time dependent mean field solution will be
products of coherent states whose labels satisfy the classical limit 
of Heinsenberg's  equations of motion as can be easily verified.
No quantum dynamical nonlocality effects appear at this level (lowest
order) unlike mean field approximations for other systems.
All quantum corrections will appear in next to leading orders as we
will show.

\paragraph{The zeroth order approximation: The mean field
approximation (MFA) and the classical limit.}
Our zeroth order approximation is defined in the following way:

\noindent
a) We consider here states of the form of products of coherent
states and a phase
\begin{equation}\label{psi}
\ket {\psi(t)}  =\exp (i\eta(t)) \ket{x(t)}
\otimes\ket{y(t)},
\end{equation}
 where the minimum uncertainty coherentes states $\ket{z(t)}$
are defined by $\ket{z(t)} = {\cal D}(z(t)) \ket{0}$
with
\begin{equation}\label{estadoscoerentes}
{\cal D}(z(t))=\left\{
\begin{array}{l}
\exp[z(t)a^\dagger-z^*(t)a]\\
\exp[\frac{\atan|z(t)|}{|z(t)|}(z(t)J_+-z^*(t)J_-)],
\end{array}
\right.
\end{equation}
for h(3) and su(2) respectively, and the fiducial state 
 $\ket{0}$ is  the Fock state $\ket{n=0}$ for h(3) and
the $J_z$ eigenstate $\ket{J,-J}$ for su(2).

\noindent 
b) Linear combinations are not allowed as initial conditions
(this circumvents problems with the superposition principle).
Also this MFA for the systems (\ref{bihamiltonian}) will leave
invariant
the manifold of the chosen set of trial functions.

At this point it is important to mention that the algebra h(3)
could be easily enlarge to include of harmonic oscillator algebra
with little effort, but the extra terms give rise to nonlocality
effects (squeezing dynamics) already at this lowest order which we
would like to avoid. Notice also that generalization to n degrees
of freedom m-linear hamiltonians ($m\leq n$) is straightforward.

Now we solve the mean field  Schr\"odinger equation 
\begin{equation}\label{MFEquation}
(i\partial_t-\bar{H})\ket{\psi(t)},
\end{equation}
where $\hbar=1$ and the mean field hamiltonian (MFH) is 
\[
\bar{H} =
 1_A\otimes H_B(x(t))+ H_A(y(t))\otimes 1_B,
\]
with  $H_B(x(t))=\bra{x(t)}H\ket{x(t)}$. 
Using the explicit form of the hamiltonian (\ref{bihamiltonian})
we find 
\begin{equation}\label{MFH}
H_A(t) = \sum_i a_i(t) A_i +f_A(t),
\end{equation}
where $a_i(t)=\alpha_i+\sum_j\gamma_{ij}\bra{y}B_j\ket{y}$
and $f_A(t)= \sum_j\beta_j \bra{y}B_j\ket{y}$.
The expression for $H_B(t)$ is completely analogous to the expression
for $H_A(t)$. Since the terms $f_A(t)$, $f_B(t)$ give rise to global
phases we neglect them. 
Due to the structure of the MFH the solutions of eq. (\ref{MFEquation}) 
will preserve the form (\ref{psi}), and their time dependence will
be completely
specified by the solutions of the following equations of motion
\begin{eqnarray}\label{clasicah3}
\dot{z}& = & -ia_0 z -ia_+  \hspace{2.8cm} (\mbox{h(3)}) \\
\label{clasicasu2}
\dot{z}& = & -ia_+ -ia_0 z +i(a_+)^* z^2  
\hspace{1cm} (\mbox{su(2)})\\
\label{accion}
\dot{\eta}_{z}& = & 
\bra{0} {\cal D}^\dagger(z(t))\left(i\partial_t -h(t)\right)
{\cal D}(z(t)) \ket{0}. 
\end{eqnarray}
Observe that the total phase $\eta(t)$ is the sum of the partial
phases $\eta_{x}(t)$ and $\eta_{y}(t)$. The nonlinearity of 
these equations arise from the self consistency of MFA. 
If the labels are scaled as $z=\sqrt{4J}Z$ for h(3),
$z=Z$ for su(2), and time as $t=t_c/(4J)$,
then the equations for $Z(t_c)$ will become independent of $J$
and correspond to the classical limit 
of Heisenberg's equations of motion.
From equation (\ref{accion})
it follows that
the phases $\eta_{z}(t)$ are generalized actions
of the coherent state $\ket{z(t)}$. 
They are of course absent from the classical limit,
but will be crucial for the quantum corrections.

Thus, we have shown that
 systems with hamiltonian (\ref{bihamiltonian}), {\em in the mean field
approximation} satisfy all the requirements we wanted: labels with
classical physical meaning, no superposition principle, minimum
quantum nonlocality effects, and
hamiltonian equations for labels which coincide with both the exact
Schr\"odinger equation and the Heisenberg equations of motion.

\paragraph{Quantum corrections.}
Now we turn to the corrections to the MFA. The exact Schr\"odinger
equation for the whole system (with the tilde indicating the 
Schr\"odinger picture)
\[
i\partial_t\tilde{\ket{\psi(t)}} 
		= (\bar{H}(t) +\tilde{\Delta}(t))\tilde{\ket{\psi(t)}},
\]
where $\tilde{\Delta}(t)=H-\bar{H}(t)$,
can be written in the MFA interaction picture (MFAIP) as (the absence of 
the tilde indicating the MFAIP)
\begin{equation}\label{MFAIP}
i\partial_t\ket{\psi(t)}   = 
	\Delta(t)\ket{\psi(t)} = 
\sum_{ij} \gamma_{ij} (A_i(t)-\mean{A_i}(t))
			      (B_j(t)-\mean{B_j}(t)) \ket{\psi(t)},
\end{equation}
again up to time dependent terms that give rise to a global phase.
Here we have defined $A(t)$ and $\mean{A}(t)$ as
\[
\begin{split}
A(t) & =\bar{U}^\dagger (t,0) A \bar{U} (t,0), \\
\mean{A}(t) & = \bra{0} {\cal D}^\dagger(x(t)) A {\cal D}(x(t))\ket{0},
\end{split}
\]
where $\bar{U}(t,0)$ is the evolution operator of the MFA, the product
of the evolution operators for each degree of freedom. The 
equations for  $B(t)$ and $\mean{B}(t)$ are completely analogous.
Equation (\ref{MFAIP}) possesses a natural expansion
\begin{equation} \label{expansion}
\ket{\psi_t} =
(1-\hspace{-0.05cm}i \int_0^t \hspace{-0.2cm} dt_1 \Delta_{t_1} 
\hspace{-0.05cm}-\hspace{-0.05cm}
 \int_0^t \hspace{-0.2cm} dt_1 \hspace{-0.15cm}
\int_0^{t_1} \hspace{-0.35cm} dt_2 
\Delta_{t_1} \Delta_{t_2} +\cdots)
\ket{I},
\end{equation}
where $\ket{I}$ is the initial state $\ket{x(0),y(0)}$.
Now, let us see that all the terms in this expansion are readily
written in terms of classical quantities. For example, the first
correction term gives 
\[
\ket{\psi(t)}=\ket{I}-iC(t)\ket{D}
\]
where C(t) is a function to be defined below and 
$\ket{D}$ is a generalized coherent state, orthogonal
 to the initial coherent state, given by
$\ket{D}=\ket{D(z(0))}= {\cal D}(z(0))\ket{1}$   
with the
reference state $\ket{1}$ given by the Fock state $\ket{n=1}$
for h(3) and by $\ket{J,-J+1}$ for su(2).
We immediately see that the first correction already introduces
all of the effects we avoided at the mean field level:
superposition of states, nonlocality and entanglement.
State $\ket{D}$
is sometimes called a doorway state. 
In order to get the first order result we made use of the 
following identity which is a result of the present group(s)
structure. In fact, it is this relation which enables one to
express all order corrections in terms of classical trajectories
and generalized coherent states. 
\begin{equation}\label{GroupRelation}
A_i D(x) = D(x)\left( \sum_k g_{ik}^A(x) A_k+k_i^A(x)\right),
\end{equation}
where $g_{ik}(x)$ are functions of $x$ specific to each
one of the groups in question.
Similar relations hold for the degree of freedom B. 
The coefficient $C(t)=\int_0^t c(t_1)dt_1$,
where 
\[
c(t) = \sigma e^{i(S_0-S_1)(t_1,0)}
\sum_{ij}\gamma_{ij}
g_{i+}^A(x(t_1))
 g_{j+}^B(y(t_1))
\]
is clearly only a function of classical trajectories and the
corresponding actions, $S_0(t,0) =\int_0^t dt_1 (\eta_{x}(t)
+\eta_{y}(t))$ and $S_1(t)=S_1[x(t)]+S_1[y(t)]$ with
\[
S_{1}[z(t)] = \int_0^t d\tau\bra{1} {\cal D}^\dagger(z(\tau))
\left(i\partial_\tau -h(\tau)\right)
{\cal D}(z(\tau)) \ket{1}.
\]
It is clear that the remaining corrections can also be written in terms
of classical trajectories (and actions), but their quantum content will
not be as transparent as in the leading correction. In fact, the
second correction can be written as having a term proportional to the
initial coherent state, but also terms proportional to 
generalized coherent states 
$\ket{x_0,y_1}$,
 $\ket{x_1,y_0}$, and 
$\ket{x_2,y_2}$, where the subindices refer to the fiducial states.

\paragraph{Sensitivity to Initial Conditions: 
A Formal Nonperturbative Result}
Classically one of the basic ingredients to define chaos is 
the high sensitivity to initial conditions. A formalization of this
condition is heavily based on the notion of distance between 
trajectories. In establishing a quantum counterpart of this condition 
it is important to introduce a quantum measure of distance
between states. A natural measure is given by the square modulus of
the scalar product.\footnote{Some proposals have been made related to
scalar product of wavefunction evolved from different hamiltonians
(not
different wavefunctions). See A. Peres, 
\textit{Phys. Rev. A} \textbf{30}, 1610 (1984);
R.A. Jalabert, and H.M. Patawski, \textit{Phys. Rev. Lett.}  \textbf{86}, 2410 (2001).} 
In what concerns our mean field approximation,
the squared modulus of the scalar product between different states 
of the manifold allows for a direct association of the distance 
between states with distances between phase space 
trajectories, since $|\bra{z_1}z_2\rangle|^2$ is given by
\begin{equation}\label{overlap}
\left\{
\begin{array}{ll}
\exp(-|z_1-z_2|^2) & \mbox{ for h(3),}\\
(1-\frac{|z_1-z_2|^2}{(1+|z_1|^2)(1+|z_2|^2)})^{2J} &\mbox{ for su(2).}
\end{array}
\right.
\end{equation}

The important quantum tool which allows us to investigate the
sensitivity
to initial conditions of the quantum dynamics and eventually make
connection to the well known classical limit is the overlap between
two time dependent states, which evolved from different initial
conditions. It is well known that for unitary evolutions the scalar
product is conserved in time. Observe that different initial states
correspond to {\em different} MFAs, since the MFA is state dependent,
due to self consistency. Thus the scalar product between two
different initial states is to be written as
\begin{eqnarray}\label{correcoes}
 \mean{x^\prime(0),y^\prime(0)|x(0),y(0)}  =
\mean{x^\prime(t),y^\prime(t)|
x(t),y(t)} 
 +\bra{x^\prime(t),y^\prime(t)} (\delta_{QC}^\prime(t))^\dagger
\ket{x(t),y(t)}
\\
\nonumber 
 +\bra{x^\prime(t),y^\prime(t)}\delta_{QC}(t)
\ket{x(t),y(t)}
 + \bra{x^\prime(t),y^\prime(t)} (\delta_{QC}^\prime(t))^\dagger
\delta_{QC}(t)
\ket{x(t),y(t)}
\end{eqnarray}
where we have written $U(t)$, the exact quantum time evolution
operator as $\bar{U}^{(\prime)}(t) (1+\delta_{QC}^{(\prime)})$,
with $\bar{U}^{(\prime)}(t)$ the MFA evolution operator
corresponding to state $\ket{x(t)^{(\prime)},y(t)^{(\prime)}}$ 
and  $1+\delta_{QC}^{(\prime)}$
the  evolution operator for the quantum corrections.

The first term on the rhs, $\mean{x^\prime(t),y^\prime(t)|
x(t),y(t)}$ contains the mean field approximation
and is given by
\[
 e^{-d(x(t)-x^\prime(t))/2} 
e^{i \Phi(x(t),x^\prime(t))}
  e^{-d(y(t)-y^\prime(t))/2} 
e^{i \Phi(y(t),y^\prime(t))},
\]
where $\Phi$ is some phase which also depends on the classical
trajectory and the corresponding group, and functions $d$ can be
determined by comparison with (\ref{overlap}).
This matrix element is proportional to the distance between the labels
which, in the present case, corresponds precisely to the classical
trajectories. Since the exact evolution preserves overlap, the sum of 
this term
with the other three of eq. (\ref{correcoes}), which contain the
quantum corrections, should be conserved in time. Observe that the
rate of change of this overlap can have two distinct origins,
dephasing and/or change of the modulus. Both changes should be
compensated by quantum corrections, but only the later one can
be unambigously connected to classical chaos, since,
if the system is classically
chaotic this distance will exponentially grow and, as a consequence, 
the overlap involving only the MFA will decrease accordingly.
This is a quantum counterpart of the fact that classically chaotic
systems exhibit high sensitivity to initial conditions. The 
corresponding quantum system will exhibit a high sensitivity 
to the initial state in what concerns 
the production
rate of non unitary quantum corrections to this overlap. 
We remark that this result is exact and independent of the
approximation
used. For the argument, however it has been crucial to separate
the mean field contribution explicitly. It should also be emphasized
that the time scale for the overlap quantum corrections is 
essentially linked to the Lyapunov exponents. However, other
observables will have different time scales, sometimes much shorter,
as
for example that for the entanglement process. Consequently, 
entanglement is not always a good measure of classical chaotic
behavior, at least for short times, unless the exponential 
separation of neighbouring classical trajectories also occurs
at very short time scales.
Should the exponential separation occur at early times, a significant increase 
in linear entropy will be noticed. In the cases the two time scales
 are very different this effect, although present, will be 
rendered less conspicuous by the time development of 
quantum correlations stemming from the other sources.
This will become clear
in the example below. 

In order to characterize the degree of entanglement we will 
calculate the idempotency defect (or linear entropy) $\delta(t)=
1-{\rm Tr_A}(\rho_A(t))^2$, where the reduced density $\rho_A(t)$
is given by $\rho_A(t)={\rm Tr_B}\ket{\psi(t)}\bra{\psi(t)}$.
Observe that this measure of entanglement does not depend on the
picture used to calculate it.
Using the expansion (\ref{expansion}) up to second order, writing
$\ket{I}=\ket{I_A}\otimes\ket{I_B}$, and calculating the
idempotency defect in the MFAIP we obtain to second order
\[
\delta(t)  = 
 4\ {\rm Re}\int_0^t \!\!dt_1\int_0^{t_1}\!\!\!\! dt_2\ 
c^*(t_1)\ c(t_2).
\]

\paragraph{Application to the classicaly chaotic maser model} 
The classically chaotic maser hamiltonian 
\[
H=\epsilon J_z +\omega a^\dagger a
+\frac{G}{\sqrt{J}} (a^\dagger J_- +a J_+)
+\frac{G^\prime}{\sqrt{J}} (a^\dagger J_+ + a J_-)
\]
belongs to the class of bilinear hamiltonians (\ref{bihamiltonian}),
where the field (atomic) degree of freedom A (B) is related to the h(3) 
(su(2)) algebra. Field coherent states are characterized by $x(t)$
while spin coherent states by $y(t)$. The MFH is of form (\ref{MFH}) 
for each degree of freedom with 
\begin{eqnarray*}
a_+ & = & -i \frac{G y +G' y^*}{\sqrt{J}},\ \ a_-=(a_+)^*
\ \ a_0=\omega,\\
a_+ & = & -i \frac{G x +G' x^*}{\sqrt{J}},\ \ a_-=(a_+)^*
\ \ a_0=\epsilon,
\end{eqnarray*}
for the field and atomic MFH respectively.
Also, the equations of motion for each degree of freedom are
equations (\ref{clasicah3}) for the field degree of freedom
and (\ref{clasicasu2}) for the atomic one. 
For this model it is a simple matter to give
an analytic expression for $c(t)$, the key ingredient for evaluation
of both the first order correction for the state, and the idempotency 
defect.
\[
c(t)=\frac{\sqrt{2}e^{i(S_{0,-J}-S_{1,-J+1})(t)}}{1+|y(t)|^2}
(G^\prime-Gy^2(t))
\]
where $S_{0,-J}$ ($S_{1,-J=1}$) is the generalized action for 
the coherent state 
with fiducial state $\ket{n=0}\otimes\ket{J,-J}$ 
($\ket{n=1}\otimes\ket{J,-J+1}$). In this case the doorway state
$\ket{D}$ is given by $\ket{D}=\ket{x_1,y_{-J+1}}$. 

\begin{figure}
\epsfig{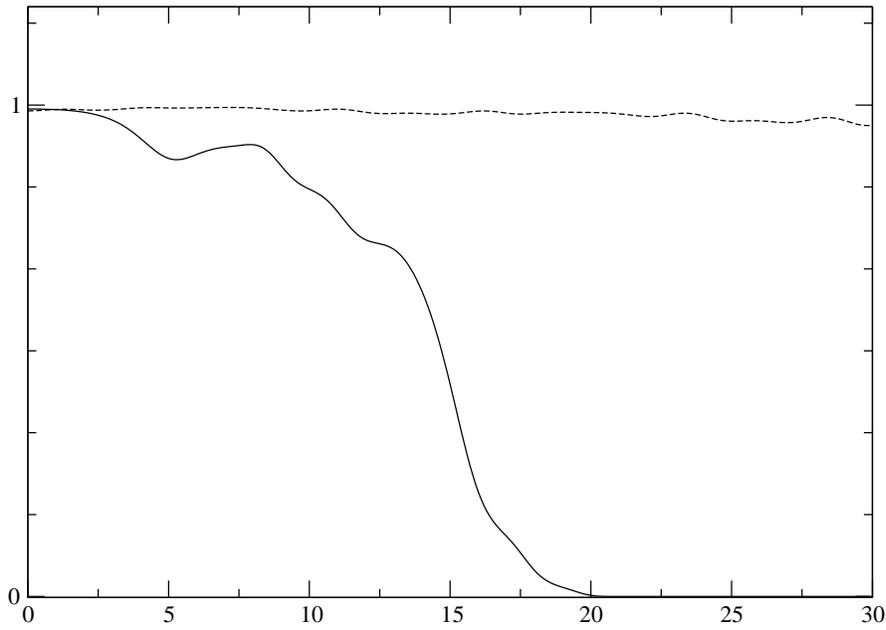}
\caption{\label{Overlap} Squared modulus of the overlap
 between two neighbouring states 
$|\mean{x_1(t),y_1(t)|x_2(t),y_2(t)}|^2$
in the MFA for mean energy E=8.5, J=9/2 in a resonant 
($\epsilon=1=\omega$)  non integrable 
case with G=0.5, $G^\prime=0.2$ for conditions in the chaotic
region (continuous lines) and regular region(dashed line). Chaotic
initial conditions $(x_1,y_1)= (5.7263433,-0.24253563)$,
 $(x_2,y_2)= ( 5.7778567,-0.26845243)$. Regular initial 
conditions  $(x_1,y_1)= ( 3.615516,0.53452248)$,
 $(x_2,y_2)= (3.68977334,0.50086791)$.}
\end{figure}

The time development of the magnitude of the overlap 
in the MFA between 
two coherent states
centered in the classically chaotic phase space region
and also the overlap between other two states chosen in the
regular region are 
shown in fig. (\ref{Overlap}), and illustrate the effect of the classically
chaotic motion on quantum dynamics is dramatic. In effect, for the
times the overlap changes appreciably, one can check that 
the classical trajectories of the chaotic region in question are also 
correspondingly well set apart. It is clear that for the magnitude 
of the overlap in MFA the time scale of correlation effects is
essentially dictated by classical dynamics. Of course, this needs 
not hold for other quantum observables. Entanglement, for example,
is a quantum property with a smaller time scale. 
The expected sudden increase in this quantity at the time the 
modulus of the overlap diminishes, disappears in the midst of
contributions of several quantum effects other than the one
related to the classical limit (see ref. \cite{Kyoko}). Our
analytical 
approximation for entanglement breaks down for times of the order 
of 2 for initial conditions both in the chaotic and regular regions.
Generalization of these results to other quantum systems is possible,
but quantum effects such as nonlocality will be already present
at the lowest order, and other analytical approximations should 
be advanced. Work along these lines is in progress.

 The authors are grateful to E.V. Passos and
M.P. Pato for fruitful discussions.
This work was partly funded by FAPESP, CNPq and PRONEX (Brazil), 
and Colciencias, DINAIN (Colombia). K.M.F.R. gratefully acknowledges 
the Instituto de F'\i sica, Universidade de S\~ao Paulo, for their
hospitality and PRONEX for partial support.


\begin{thebibliography}{9}
\bibitem{entanglement} W.H. Zurek.\textit{Phys. Rev. D} 
\textbf{24}, 1516 (1981). For
recent reviews J.M. Raimond \textit{et.al.}, \textit{Rev. Mod. Phys.}
\textbf{73}, 565 (2001);
A. Zeilinger, \textit{Rev. Mod. Phys.}  \textbf{71}, S281 (1999) 
and references there in.
\bibitem{Bohigas} O. Bohigas {\textit et.al.},
\textit{Phys. Rev. Lett.} 
 \textbf{52}, 1 (1984).
\bibitem{Gutzwiller} M.C. Gutzwiller. {\textit J. Math. Phys.} 
\textbf{12}, 343 (1971).
\bibitem{Zurek} W.H. Zurek, and J.P. Paz, Physica (Amsterdam) 
\textbf{83D}, 300 (1995)
\bibitem{Kyoko} K. Furuya \textit{et.al.}, \textit{Phys. Rev. Lett.} 
 \textbf{80}, 5524
(1998). 
\bibitem{Miller} P.A. Miller and S. Sarkar, \textit{Phys. Rev. E} 
\textbf{60}, 1542 (1999).

\end{thebibliography}
\end{document}